# Scoping Review: Mental Health XR Games at ISMAR, IEEEVR, & TVCG

**Cassidy R. Nelson**[1]
Division of Games
Unviersity of Utah*

**ABSTRACT**

Extended reality serious games for mental health are a promising research avenue to address the accessibility gap in mental health treatment by bringing therapy to patients in their homes, offering highly adaptable and immersive yet safe therapy opportunities, and increasing motivation and engagement with therapeutic exercises. However, the sensitive use case of mental health demands thoughtful integration with mental health concepts and a comprehensive understanding of prior literature. This paper presents a scoping literature review of the ISMAR, IEEEVR, and TVCG communities to assess the contributions of the XR community to the mental health serious game domain and explore potential weaknesses and strengths for future work by XR researchers. To this end, this review identified 204 possibly relevant articles in the XR community and fully evaluated 6 XR serious games for mental health. This relatively small number of articles for final inclusion suggests that XR mental health serious games are largely underexplored by the XR community (or not reported within the XR community). There is value in exploring the existing literature space as it is. Thus, this paper evaluates these six papers in terms of game elements and underlying psychological foundations, and discuss future directions for XR researchers in this wide-open research space within our community.

**Index terms**: Extended reality, serious games, mental health, human-computer interaction, systematic review

## 1 INTRODUCTION

About half of the global population will develop at least one mental health disorder by age 75 [1]. This estimate includes *any mental illness* (AMI) of mild to moderate intensity, as well as *serious mental illness* (SMI), consisting of severe mental illness resulting in disability [2]. AMI and SMI sufferers experience the adverse sequela of their mental illnesses (MIs) personally, socially, familially, professionally, and economically [3], with negative impacts also experienced by loved ones, colleagues, and more. Despite global acknowledgement of accessibility issues and several advancements in types of mental healthcare [4], problems with accessibility persist [5]. Access is often considered in terms of affordability, the number of providers, the distance between providers and patients, and access to relevant resources, such as medications. However, realized access to care is also impacted by social stigma, patient motivation, and the difficulty/intensity of the care itself [6]. Extended reality (XR) technologies have the potential to bridge some of these access gaps [7], [8] such as literal access to care in a mental health desert. Serious games can provide additional bridging by increasing patient motivation [9] to actively engage in care.

XR technologies, such as augmented and virtual reality, offer novel, mobile (home-based), multi-sensory, graded-exposure, and interactive therapy opportunities [10], while providing therapeutically valuable, precise, and controllable challenges that are customizable to each patient [10]. XR can facilitate both treatment and diagnosis of mental health disorders [11] by creating safe simulations and integrating affective measures. Such simulations can include basic exposure therapy [12] or nuanced virtual agent conversations (sometimes powered by AI) [13]. A 2024 systematic review revealed that XR therapy can lead to a significant reduction in symptoms of anxiety and depression [14]. However, despite their clinical value, some therapies can be repetitive and boring [6], causing patient attrition.

*Originally Ohio University at time of publication

Serious games, or games with a primary purpose other than entertainment, can both motivate continued (and perhaps more earnest) engagement with therapy [6], have been shown to moderately improve mental health symptoms [15] and are a quickly expanding area of mental health treatment research [16]. This is because serious games can create a more holistic and enriched therapeutic experience through narratives, non-player character interactions, quests, and other elements. However, most serious games to date are facilitated via desktop computer games [17]. While valuable, these games miss opportunities to integrate therapy into the user's environment.

XR serious games are identified as a promising technology for the future of meaningful human use-cases like healthcare, learning, and mental health (and a grand challenge for human-computer interaction) [7], [8] due to the distinct benefits unique to this specific merging of experiential strategies (like real-time integrated XR feedback and comprehensive measurement opportunities [18]). Moreover, the two combined can have a resonant effect and further improve motivation [18], immersion, and more enriching experiences that would otherwise be infeasible [19].

Successful design of XR serious games for mental health (MHXRGs) will demand interdisciplinary, thoughtful integration of therapeutic theory [16], [20]. Furthermore, the complex and sensitive nature of MHXRGs requires a thorough understanding of prior work. This paper is inspired by five successful ISMAR workshops to date, focused on XR research for mental health (MARMH), indicating a potential emerging interest trend within this research community. While individual prongs into mental health research are meaningful, a systematic literature review can provide a holistic view of the research domain and provide insights that can guide future work to better realize the unique benefits of XR, games, and XR games for MH [18].

To systematically investigate the use of MHXRGs, this scoping review focuses on research disseminated through premier A* XR sibling academic conferences, The International Symposium on Mixed and Augmented Reality (ISMAR) and the IEEE Conference on Virtual Reality and 3D User Interfaces (IEEE VR), as well as their shared journal publication venue IEEE Transactions on Visualization and Computer Graphics (TVCG). Limiting a literature review to specific research communities has been employed before [21] as a strategy to identify weaknesses and highlight the strengths of particular research communities. In fact, community-limited review has already been done on VR serious games for education within the learning community [21]. By initially limiting the scope of this review to the XR community, we can identify emerging trends and potential gaps for the XR community to focus on in future work.

### 1.1 Contributions of this Work

This work offers an exploration of the existing MHXRG literature space within premier, sibling XR conference venues and their shared journal venue. This paper provides an overview of the literature space, identify strategies for future MHXRSGs, evaluate the game experience itself, the XR interactions, the underlying mental health theories, and finally discuss future considerations for MHXRSG researchers in the XR community. While six papers may seem too small to conduct a meaningful review, scoping reviews are inherently exploratory and seek to evaluate the literature as it is. There is no formal minimum requirement for reviews [22] and protocols exist for even zero article reviews (e.g., reviews that revealed no relevant literature based on the research question) [23]. The possibility of zero-review papers, combined with the exploratory nature of this scoping review, suggests that six papers provide more than enough for a meaningful discussion of the current literature space.

### 1.2 Related Work

Lau et al. review the accessibility, feasibility, and effectiveness of serious games of computer-based desktop games [15]. While their review does not include XR games, they do note that serious games may be effective in reducing symptoms of mental health disorders. Luong et al. offer a literature review of VR for mental health via an evaluation of affective and cognitive VR broadly (not focused on games). While their review captures and briefly touches on VR games, it does not delve into specific details about them beyond noting that games can be used to successfully facilitate emotional scenarios [24]. Heidarikohol & Borst offer a survey of non-immersive and XR games for emotional intelligence [25]. However, their review focuses on classifying which emotional intelligence models are used in the literature, rather than on the gameplay experience itself or how XR enhances that experience. To our knowledge, no other reviews have been conducted that explore serious games for mental health in the XR domain. Our prior systematic reviews focus on XR serious games, one for physical rehabilitation [18] and one for learning [20]. Moreover, our prior reviews inspired the evaluation strategy for this paper. E.g., XR games themselves are explored and how therapy (instead of learning or exercise) concepts are embedded within them. Serious games for vulnerable domains, like mental health, demand sensitive consideration of foundational principles of their intended use case [20]. Thus, this work leverages their framing and evaluation strategy, but for mental health.

---

[1]email: cassidy.nelson@utah.edu

## 2 METHODS

This scoping review leverages the Preferred Reporting Items for Systematic Reviews and Meta-Analysis (PRISMA-ScR) [26], [27] scoping review framework to ensure the review is comprehensively and transparently reported. A scoping review was opted for over a systematic review, as this is an exploratory analysis aimed at mapping key concepts within a specific research community, identifying knowledge gaps, and summarizing evidence. Expanding beyond this specific research community in future work
could yield a more comprehensive systematic literature review.

ISMAR and IEEEVR are sibling A* conferences with their own conference proceedings, workshop proceedings, and a shared host journal titled IEEE Transactions on Visualization and Computer Graphics (TVCG). Thus, this literature review encompasses all ISMAR conference and workshop proceedings, IEEEVR conference and workshop proceedings, and all TVCG publications.
To ensure IEEE Xplore only searched within these specific publication venues, a search was conducted within each parent publication page. When the filter *"Search within Publication," is clicked,* the unique unique Parent Publication Number for each venue is reported:
- TVCG Journal – 2945
- ISMAR Conference – 1000465
- ISMAR Workshops – 1810084
- IEEEVR Conference – 1000791
- IEEEVR Workshops - 1836626

Thus, five searches were run with each unique parent publication number using the following search string: *(gam\* AND "mental health" OR therap\*) NOT physical*

We used a wildcard (*) for gam* to indicate game, gamification, and games. Note that using gam* will also capture serious games. To capture mental health, the search terms "mental health" and therap* were used to reduce risk that relevant papers exploring therapy but not using the term MH were incorrectly excluded. Finally, the term 'physical' was excluded to filter out papers related to physical therapy. Each of these venues focuses on XR technology (except TVCG), so XR-related language was not included in the search terms. As the XR community struggles with linguistic chaos [20], [28], it was ideal to manually sort out non-XR papers from TVCG instead of entering all potential relevant search terms like "virtual reality, mixed reality, augmented reality, extended reality, augmented virtuality," etc., to maintain consistency with all database searches. This search string (coupled with each Parent Publication Number) gave us the following results:
- TVCG journal – 40 results
- ISMAR Conference – 12 results
- ISMAR Adjunct – 44 results
- IEEEVR Conference – 40 results
- IEEEVR Adjunct – 68 results
- TOTAL: 204 results for consideration

Papers were assessed to determine whether they explored an 1) XR  2) game 3) for mental health. Those terms are operationalized as follows:

1) XR includes augmented reality, virtual reality, and augmented virtuality [29] that is registered in 3D, interactable, and context aware [30]. 'XR' does not include standard desktop computer-based interfaces.

2) In the literature, 'games' as a term encapsulates both gamification and serious games despite their distinct nature [20]. Serious games are games with a primary purpose other than entertainment [15], creating a completely new task experience, whereas gamification adds game-like elements to a standard task [20]. This review also required that papers describe their games in enough detail to ensure they are truly games. Games must comprise game elements (such as points, timers, and health bars), rules to define game parameters, objectives for players to achieve, and a certainegree of challenge [20]. Papers that label themselves as a 'game' but only describe a standard task completed within VR are not included by this review (e.g., if the same task outside of the novelty of VR would not be considered a game, it was excluded).

3) Mental health as a research term is often conflated with cognitive rehabilitation. Mental health research, as defined by the American Psychological Association, is associated with emotional well-being, behavioral adaptations, reducing anxiety or disabling symptoms, and fostering a capacity to establish constructive relationships with ordinary demands and life stressors [31] for emotional, mood, and behavioral disorders like depression, anxiety, bipolar disorder, etc. An example of this is exposure therapy, which helps patients experience gradual increases in triggering stimuli in a safe environment. Cognitive rehabilitation, however, is defined by the American Psychological Association as addressing issues with cognitive processing resulting from chronic illness like dementia or brain injury like a stroke or concussion [32]. Research in cognitive rehabilitation centers on mental (cognitive) abilities such as joint attention, working memory, and hand-eye coordination. There can be overlap in these research domains, like exploring cognitive skill building to mitigate the impacts of schizophrenia, or perhaps assessing how improved joint attention in people with autism can also lessen social anxiety. Such overlaps are included in this review, whereas papers focused solely on cognitive skill building with no connection to mental health concepts (such as hand-eye coordination) are excluded from this review.

Paper search concluded June 24, 2025. 72 papers were eliminated for not including or focusing on mental health, 125 for not being a game, and 1 for not being XR, leaving behind 6 for deeper review (see Fig. 1). The greatest reason for exclusion was that the paper did not describe, evaluate, or focus on a game. The final selection of articles includes 1 TVCG paper, 2 IEEEVR conference papers, and 3 IEEEVR adjunct proceedings.

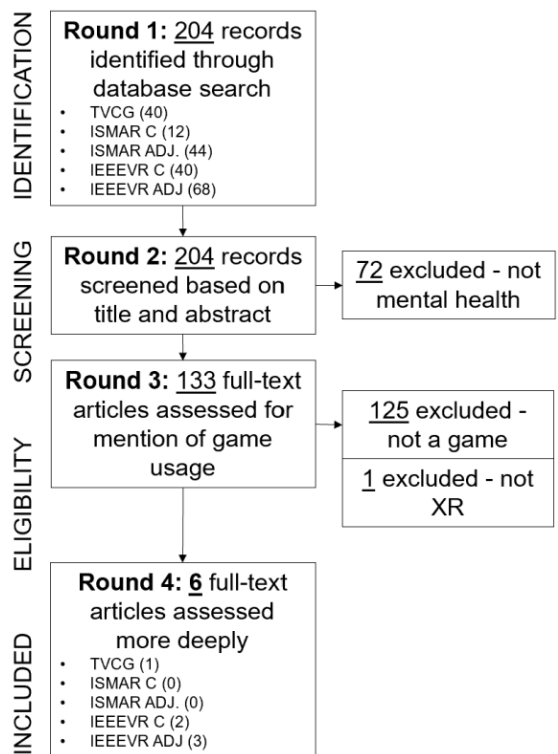

Figure 1: Article filtration process following PRISMA

# 3 RESULTS

## 3.1 Trends

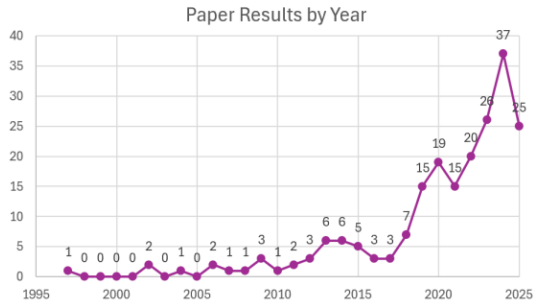

Figure 2: Publication Trends

The initial 204 results spanned from 1997 to 2025, with a notable increase beginning in 2008. This trend is evident in Figure 2, and it is hypothesized that by the end of 2025, there will be more papers than in 2024. The final six included works span 2014-2025. Figure 1 shows a relatively consistent number of articles between databases, with the ISMAR conference proceedings being a standout low and the IEEEVR adjunct proceedings a standout high.

Figure 1 outlines how many papers were eliminated for either being not about mental health, not about a game, or not using XR. Note that once a paper was excluded for one of these reasons, other relevant parameters were not tracked (i.e., it is possible that a larger % of papers were not games, etc.). 61% of papers were excluded for not being a game, 35% for not being a mental health application, and .49% for not using XR.

Table 1 outlines the types of XR used and the intended mental health use case of the game. Note that none of the final included papers used augmented reality, while five papers used head-mounted VR devices.

Table 1: Type of XR and Mental Health Application

| Cite | Venue | Year | XR Type | Use Case |
|---|---|---|---|---|
| [33] | TVCG | 2014 | VR Cave | Sport Psychology - Performance Anxiety |
| [6] | VR CONF | 2019 | VR HTC VIVE | Alcohol Use Disorder Therapy |
| [34] | VR CONF | 2021 | VR Quest | General non-clinical Anxiety |
| [35] | VR ADJ | 2021 | VR Quest | Major Depressive Disorder |
| [12] | VR ADJ | 2022 | VR (unreported HMD) | Agoraphobia |
| [36] | VR ADJ | 2025 | VR (unreported HMD) | Gambling Disorder |

Colleagues (Peck et al.,) in the XR research community have highlighted a concerning lack of equity in the VR literature [36], particular in the context of vulnerable user populations [18]. Thus, this work evaluates participant demographics (Table 2). Four papers evaluated their games with actual participants. 94 reported participants were captured by this review with an age range of 17-48 and an approximately 43/57% female/male participant split, respectively. Unfortunately, one study did not report the demographics of their user study or the number of participants. Two studies did not evaluate their games but did work with the relevant stakeholders in the development of their game. No papers report ethnicity but this review did capture the country of origin for each paper based on author affiliations. The represented countries are the USA twice, Germany, Canada, Ireland, and Italy.

In Table 2, "Intended User" refers to the purported target demographic for the MHXRG. Gen. pop. refers to the general population. Both papers targeting the general population are evaluating non-clinical anxiety. PW refers to "people with…" a disorder.

Table 2: Participant Information

| Cite | N | Demographics | Intended User | Participant Alignment |
|---|---|---|---|---|
| [33] | 25 | Age 22-32, 36% female, USA | Gen. Pop. Anxiety (Non clinical) | Alignged |
| [6] | 13 | Age 22-35, 31% female, Germany | People with (PW) Alcohol Use Disorder | Unaligned – Gen. Pop. |
| [34] | 56 | Age 17-48, 54% female, Canada | Gen. Pop. With Anxiety (non clinical) | Aligned |
| [35] | 2 clinicians, 2 patients (to inform game dev) | Unreported, USA | PW Major Depressive Disorder | N/A Eval, Aligned in Dev |
| [12] | 0 Eval (34 pre-survey to inform game dev) | Unreported, Ireland (online survey unknown) | PW Agoraphobia | N/A Eval, Aligned in Pre-Survey Phase |
| [36] | Yes – Unreported | Unreported, Italy | PW Gambling Disorder | Unreported |

Two papers focus on addiction (alcohol or gambling), one on major depressive disorder, and another on agoraphobia. "Participant Alignment" refers to whether the captured participant pool matches the intended final user (e.g., if the XRMHG is intended for people with agoraphobia, were the participants people with agoraphobia). The only papers with aligned participants were papers focusing on the general population.

### 3.2 Games Described

Given the limited scope of MHXRGs in our research community, it is prudent to describe the existing literature in more detail. In so doing, this paper consolidates examples for future scholars and evaluates existing game design, study methodology, underlying psychological grounding, and ultimate research findings.

#### 3.2.1 <u>Virtual Goalkeeper</u>

Stinson & Bowman [33] offer a feasibility assessment of training athletes for high-pressure performance anxiety in VR. This work takes an existing game concept in the physical world and uses the adaptability and customizability of VR control variables that may impact competition anxiety to assess whether a VR version of a physical game can induce anxiety. If so, a VR version of physical sports games could serve as an exposure (inoculation) against performance stress.

Goal
Evaluate a simulated soccer goalie task to see whether performance anxiety can be induced. To this end, the study evaluates three independent variables at high and low levels: Known anxiety triggers, field of regard, and simulation fidelity. *Known anxiety triggers* were lack of control, unpredictability, and negative feedback. In 'low' anxiety trigger conditions, the simulation had no negative feedback, the kicks were predictable, and participant performance reflected their actual success. In a 'high' anxiety trigger, a deliberate lack of control is introduced by saying participants 'missed' even if, technically, they had succeeded. Unpredictability introduced an unpredictable delay before each kick. Negative feedback was in the form of discouraging messages on the screen in red letters. *Field*

*of Regard* referred to the degrees of horizontal visibility, ranging from 90 (low) to 270 (high). *Simulation fidelity* had to do with immersive elements. In the low condition, the task took place in a wide open soccer field with only blue sky and grass, the opposing kicker, and a floating score above the kicker. Sounds included a referee whistle and the ball kick noise. In the high condition, the environment is a soccer stadium full of audience members, other soccer players in the field, more kick animations, and additional crowd noises like ambient chatter, cheering/booing.

Game Design
Soccer goalie simulation. Participants had to effectively block an incoming ball with varying parameters, including the presence of an audience, negative feedback via written words that compelled participants to do better, and (unknown to participants) some deliberate smudging of the game to make players 'miss' a ball they otherwise would have caught. Points could be won.

Psychological Theory
This work explicitly references sport psychology, competitive anxiety, and how anxiety relates to decision making and performance via the Hebbian take on the Yerkes/Dodson law of arousal vs performance. Moreover, this work also touches on the VR exposure therapy (VRET) and its prior success as a graded-exposure methodology to reduce stress. Several measures were based on psychological theory to assess somatic and cognitive anxiety, as well as confidence.

Measures
Galvanic skin response, heart rate, heart rate variability, state-trait anxiety inventory for cognitive and somatic anxiety (STICSA), Competitive State Anxiety Inventory 2 Revised (CSAI-2R), Performance (Save %, Reaction Time), interviews

Findings
Anxiety can be induced in VR sports subjectively and physiologically, though it is unclear how much physiological measures were impacted by physical exertion. The known anxiety triggers (negative feedback, lack of control, unpredictability) have a direct relationship with anxiety scores. Lower field of regard created higher anxiety and lower confidence (hypothesized to be a tunnelling effect on performance) with lower simulation fidelity. High simulation fidelity increased anxiety. There were no clear anxiety/performance patterns (thus no exhibited Hebbian curve). Participants with higher trait anxiety were more anxious generally, but not significantly so. Prior soccer goal keeping experience had significant interaction with simulation fidelity (low experienced people had higher anxiety with a more severe uptick in the higher fidelity condition), and anxiety decreased confidence in the low experience group.

### 3.2.2 Shopping to Avoid Alcohol

Mostajeran et al. [6] take a standard grocery shopping experience and convert it to a VR game. They deliberately selected grocery shopping as this is a higher risk environment for recovering alcoholics. Their game embeds cue avoidance therapy (CET) and approach avoidance training (AAT) within the shopping experience. Thus, this work provides VRET graded exposure and a more motivating way to engage in repetitive cue avoidance therapy (explained below).

Goal
Evaluate three games for their ability to render Cue Exposure Therapy (CET) and Approach Avoidance Training (AAT). Compare game-based approach avoidance training vs non-game VR based AAT.

Game Design
THREE GAMES; Overall shopping game: Buy classes of items on a shopping list (Fruit, hot drink, etc.) to allow freedom of choice. Correct item selection = green flashing cart and positive auditory feedback. Wrong item = red shopping cart flash and negative auditory feedback. Alcoholic beverages are located randomly throughout the store. However, participants had to earn money for their groceries with mini games that provided cue exposure therapy and approach avoidance training. AAT game: Sort non-alcoholic beverages into a cart and throw away alcoholic beverages. Reward of .5 euro plus positive auditory and visual (green) feedback for correct, losing .5 euro and negative auditory and visual feedback (red). Time limits included. CET game: This game offered longer exposure to alcoholic beverages. The task was unloading bottles and shelving them in designated locations. There was another time limit, but no penalty for wrong answers. Only an increase of .5 euros for each bottle placed. The non gamified AAT was simply sorting alcohol and non-alcohol in a 'home' VR environment with no timer, points, or positive/negative feedback.

<u>Psychological Theory</u>
Embedded established exposure therapies to reduce craving when exposed to alcohol and to increase reflexive ability to avoid alcohol.

<u>Measures</u>
Focused on broad UX of the games. System usability scale (SUS), AttrakDiff (UX questionnaire evaluating pragmatic quality, hedonic quality, and attractiveness), NASA TLX, subjective (homebrewed) enjoyment/motivation/realism questionnaire, logged performance accuracy.

<u>Findings</u>
WG (whole overarching game containing the gamified CET, gamified AAT, and overall shopping experience), GAAT (gamified AAT), NAAT (non gamified AAT). Good SUS scores were broadly reported (all over 86); GAAT was considered more creative, challenging, novel, and of greater value than the non-game version. More stimulated by the game version. The game version is not more attractive than the non-game version. TLX: The game version had higher mental, physical, and temporal demands, with no significant difference in frustration or performance. In objective performance, participants were significantly more accurate in the game version. Participants reported enjoying the increased challenge of the game version, as the other non-game version was found to be boring.

### 3.2.3 <u>Beat Saber for Anxiety</u>

Hawes & Arya [34] created a Cyclical Priming Methodology integrated within an established Experiential Learning Theory model. This study seeks to explore the PEP (preparatory experience priming) step of that model. For example, Hawes & Arya aim to determine whether different types of priming before a concrete learning experience can reduce anxiety and thereby improve cognitive bandwidth. The justification for VR itself stems from prior work showing that VR can reduce anxiety. Thus, the actual unique VR implications of this XRMHG are not well described, justified, or explored.

<u>Goal</u>
This paper compares a commercial VR game (Beat Saber) and a separate commercial non-game VR meditation app (Calm) for their ability to serve as a priming experience before a cognitively demanding task.

<u>Game Design</u>
Standard Beat Saber demo experience played twice with no specific changes made by the research team. Participants must hit the incoming bricks with a left or right lightsaber (denoted via a distinct color for each hand) in a specific direction based on the block color (to denote which saber to use) and an arrow on the block to denote which direction to slash.

<u>Psychological Theory</u>
Authors considered Experiential Learning Theory, Positive Affect Priming, theories of anxiety being based on cognitive biases and scarcity mindset, cognitive priming, and motivational mindset. Authors also considered VRET and cognitive based therapy (CBT) in the development of their Cyclical Priming Methodology.

<u>Measures</u>
Univ of Cali Matrix Reasoning Test (UCMRT), State Trait Anxiety Inventory (STAI), Homebrewed UX survey

<u>Findings</u>
VR game priming yielded significant improvements in cognitive test scores while VR meditation did not. Both VR experiences significantly improved anxiety, but there was no significant difference between the VR game and the VR meditation experience.

### 3.2.4 Schwer

Li & Luo [35] created a new serious game for people with depression and loved ones of people with depression by working with two clinicians and two people with clinical depression. VR was leveraged due to prior work showing that the improved immersion and embodiment in VR also improved the efficacy of empathy-building experiences.

Goal

This paper describes (but does not evaluate) a game intended to teach people with major depression how to reach out for help and to teach loved ones of people with depression about the experience so they can build empathy. The game serves both as an empathy training experience and as a de-stigmatization therapy.

Game Design

Schwer is a first-person adventure maze puzzle game with a narrative and levels. Overall game progress is indicated by ambient light and water level in the world (darker, more water = lower achieved game progression and deeper depression). Player stamina (teleport range) is indicated via game progress and is tied to a bar of emotional valences ranging from "anxiety" to "numbness" (participants want to keep the bar in the middle of the range). NPCs provide quests and advice for keeping the stamina bar in the middle. There is a brain icon indicating player attributes in stamina, intelligence, interaction ability, and emotional capacity. The level of these goes up or down based on player choice. Quests require participants to solve puzzles and collect supplies. There are several NPCs that provide feedback: The Wise is a psychologist, The Sage is family or friends, and The Fool represents people who do not understand depression. Players want to follow the guidance of The Wise and The Sage while avoiding or ignoring The Fool. The Fool provides misleading or insulting guidance. The game has challenges, color feedback, rewards, and positive reinforcement.

Psychological Theory

This paper leverages prior work on empathy training and separate work using VR as an "empathy machine."

Measures, Findings,

This game was unmeasured at the time of publication, but Non-Player Characters (NPCs) have embedded in them the Form of Self-Criticizing/Attacking & Self Reassuring Scale (FSCRS). Other future evaluation strategies were not outlined.

### 3.2.5 Agoraphobia Virtual Walk

Barnett et al. [12] describe the creation of a new serious graded exposure VRET game as a therapeutic strategy for people with agoraphobia. The game design was informed via an online survey of 34 people with agoraphobia asking them what affects their anxiety the most.

Goal

Describe (not evaluate) a game intended to help people with agoraphobia venture into a virtual outside world (graded exposure) through gamified challenges within virtual outdoor environments.

Game Design

Goals via tasks, gradual increasing difficulty, the game gradually pushes people further out of their comfort zone (further from the virtual house). The game is not described in more detail.

Psychological Theory

This paper references cognitive behavioral therapy (CBT) and exposure therapy.

Measures, Findings

Not evaluated, no findings, no reported future measures.

### 3.2.6 P.E.T.R.A. Amusement Park

Papapicco et al. [36] created a virtual amusement park environment comprised of several gambling games or experiences. The goal was to create a realistic test bed environment to simulate factors that increase gambling so future work can use a circumplex model of emotion to track affective states in real time to understand impulsive decision making. Future work will capture human behavior, emotional, and physiological responses in gambling contexts. VR was selected to improve immersion but is not described further.

<ins>Goal</ins>

Describe (and only evaluate for basic SUS usability) a Persuasive Environment for Tracking and Regulating Arousal and Valence (PETRA) in VR that can simulate psychological and environmental factors that increase gambling.

<ins>Game Design</ins>

Gambling parameters like ride speed (because it is an amusement park with roller coasters), sensory feedback (auditory and visual), and reward probabilities are all alterable. The 'park' contains six experiences - Not all are games. *Bouncing Odds* and *Illusion Wheel* are games that explore reward anticipation and disappointment. *Dicey Choices* explores the illusion of control. The games themselves are not described in more detail. Non games include *High Stakes* and *Lucky Fall* rides that heighten arousal to simulate the excitement of gambling. *Wagering Wheel* (like a Ferris Wheel) provides a calming experience to give an opportunity for introspection and emotional regulation.

<ins>Psychological Theory</ins>

The authors reference the Circumplex Model of Emotion that the park is built around and cognitive behavioral therapy.

<ins>Measures</ins>

System Usability Scale for general UX testing, indicated that future work will use Circumplex Model of Emotion, EEG, and Galvonic Skin Response

<ins>Findings</ins>

Good sus scores (77.8) with high immersion as measured via subjective verbal feedback from participants during play.

## 4 DISCUSSION

Our paper evaluated 6 (from 204 initial) records for a scoping review of XR mental health serious games (XRMHGs). This final inclusion of six articles reveales ample opportunity for the ISMAR, IEEEVR, and TVCG community to explore XRMHGs. Each of the included works are outlined in detail (Section 3.2) and this section discusses the implications of the holistic literature body.

### 4.1 Be Creative and Strategic for Maximum Effect

The first research domain that comes to mind when considering XRMHGs is likely a heavily applied therapeutic context, such as VRET. However, this review revealed that while XRMHGs were used to facilitate therapeutic interventions, they could also be a strategy to test components of a psychological theory, or provide a clever testbed for deeper psychological phenomena. Strategically created XRMHGs can test therapeutic benefits and ascertain underlying psychological phenomena simultaneously.

XRMHGs can be off-the-shelf commercial games, recreations of existing physical games but in XR, or more holistically developed from-the-ground-up experiences. Games captured by this review ranged from simple (block a soccer ball) to complex (a narrative puzzle game about depression), displaying the range of potential manifestations for XRMHGs. This range of complexity gives ample opportunity for future researchers to explore XRMHGs for simultaneous therapeutic benefit and basic phenomena evaluations. To maximize the value of XRMHG intensive development and the use of vulnerable participants for scientific purposes, XRMHGs should endeavor to make the most of their game (and experiment).

### 4.2 Thoroughly Consider the Game and XR Experience

Our prior work has noted that commonly, XR serious game papers do not describe their games in much (if any) detail at all [18], [20]. Due to this lack of detail, it is also often unclear what XR specific gains, affordances, and interactions are embedded within the experience [37]. Both of these trends remain largely true for this literature review, and our the set of included work had a mix of heavily detailed and lightly detailed XR game papers.

The use of serious games was strategically leveraged across studies to support motivation, immersion, and distraction, as well as to provide a play-based mechanism for engaging with difficult psychological topics or eliciting specific psychological phenomena for study. Games comprised several established game elements [20] like NPCs, points, puzzles, collection, and narratives, with more comprehensive experiences comprising a higher amount of game elements.

In comparison, extended reality (XR) technologies contributed additional benefits such as graded immersion, safer therapeutic environments, and enhanced ecological validity in simulating real-world experiences. While the integration of game elements was often well-justified and explained, the unique contributions of XR (or how XR was tied to the experience) were often underarticulated or superficially addressed. For example, *Beat Saber for Anxiety* lightly pointed to immersion helping reduce stress and only one paper (*Schwer*) explored an XR specific interaction via game fatigue for teleport distance.

Future work should consider other opportunities to integrate XR specific interactions into the MHXRG holistic experience, as well as more explicitly justify the use of XR by grounding its components in prior empirical or theoretical literature, ensuring that its inclusion is not merely novel but meaningfully aligned with therapeutic or research goals. No augmented reality MHXRGs were found for review, indicating a gap in XR modality for MHXRGs. Moreover, MHXRGs should be considerate of how game elements themselves inform the intended therapeutic/research outcomes, as well as integrate with XR specific capacities / interactions to create new game experiences.

### 4.3 Ground in Psychology: From tasks to Foundations

Similarly to 4.1.1., the first idea to integrate psychology into XRMHGs is likely to gamify existing psychological tasks. The most leveraged psychological / therapeutic underpinning was cognitive behavioral therapy and exposure therapy. However, the realized infusion of psychological concepts was diverse. *Shopping to Avoid Alcohol* provided a thorough theraputic grounding for each and every game experience by gamifing established theraputic activities [6], *PETRA* underpinned their entire game suite with a foundational psychological theory (Circumplex Model of Emotion) [36], *Schwer* [35] created a holistic game experience based on patient experiences of depression, *Beat Saber for Anxiety* did not implement any psychological phenomena in the game but instead inserted the game in a specific part of a psychological interaction framework (Experiential Learning Theory) [34], and *Virtual Goalkeeper* used a game to elicit specific emotional responses [33]. This again outlines the range of possible strategies to implement games into mental health research. Future work should thoroughly ground or tie its XRMHG to mental health theories, therapeutic strategies, and measurement strategies, and these details should be provided in the paper, tying them to game elements and XR interactions.

### 4.4 Evaluate Rigorously (and Carefully)

MHXRGs are inherently intertwined in a vulnerable and sensitive application space, where highly vulnerable users are involved. Some MHXRGs (like those for addiction) could even be quite triggering and cause regression if not carefully implemented. Future work should consider integrating clinicians specializing in their intended user population to both oversee participant safety and to provide appropriate, theoretically based evaluation strategies like the state-trait anxiety inventory for cognitive and somatic anxiety (STICSA). Moreover, future work should embed themselves in relevant psychological literature to assess whether other XRMHGs exist for that specific clinical domain outside of the XR community.

Only two papers in this review successfully evaluated their game with the appropriate target user type and it is hypothesized to be non-trivial that those are also the only papers targeting the general population. Future work on XRMHGs should thoughtfully and rigorously consider how to access their target population before development begins. Moreover, researchers should be sensitive to gender representation, as the studies captured by this review had an unjustified, unbalanced participant split of 43/57 % female to male. E.g., it is possible and likely for some mental health disorders to distinctly impact one gender more than another [5], but this should be reported and indicated with citations in the paper.

Overall, the XRMHGs were found to be effective at eliciting specific emotional phenomena and motivating engagement with therapeutic tasks. Measurement strategies spanned from purely subjective self-reported UX surveys to psychologically validated surveys and physiological data. Using several concurrent measurement strategies (like *Virtual Goalkeeper*) can provide redundancy, validity, and unearth new relationships between physiological measures and psychological phenomena.

### 4.5 Too much detail to provide about the XR game?

Interestingly, half of the papers did not include an evaluation with participants. The lack of evaluation is not that strange given the volume of workshop papers included in the final review. However, let us note an interesting potential strategy for the XRMHG space. XRMHGs (or any XR serious games) demand a high level of interdisciplinary consideration to maximize efficacy and value. This, plus the slow development cycle of a comprehensive XR serious game, likely contributes to the relatively small number of XRMHGs captured by this scoping review. Such projects can be a large and slow undertaking for the academic publication cycle. Perhaps future developers should consider publishing detailed papers about their XR games as workshop papers. This gives developers an opportunity to get community interdisciplinary feedback, provides an additional publication for the amount of effort required to create an XR serious game, and creates an avenue for any future papers conducting a formative analysis for a conference or journal level submission to save precious space by not needing to describe the games or XR experiences in detail as the prior workshop paper can be cited.

### 4.6 Use Semantic Precision

Most of the papers eliminated from the review were due to linquistic chaos [28] surrounding the terms *game* and *mental health,* respectively. Prior review has identified the phenomenon of alleged "game" papers that boil down to standard tasks moved to VR, not comprising gamification or a full game experience [20], and and this phenomenon impacted this work as well. While it was anticipated that *therapy* would overload between mental health and physical therapy concepts (Section 2), it was not anticipated that *therapy* would also overload so severely on cognitive rehabilitation therapies. Compellingly, only three of the 133 papers included for full text review actually used 'cognitive rehabilitation' as a term or tag. Thus, the literature body has a tendency to truncate *mental health* applications and *cognitive rehabilitation* applications under the shared term *therapy*. Future XRMHGs and XR serious games for cognitive rehabilitation should ensure they use the full term "mental health therapy" or "cognitive rehabilitative therapy" to ease some of this linguistic chaos in the literature. More specifically, future work should ensure that "mental health therapy" applications actually center on the American Psychological Association's definition of *mental health* (fostering emotional well being) and not only on improving a cognitive facet of a condition (like building attention skills in ADHD) [38]. Cognitive rehabilitation papers interested in mental health should consider how the cognitive condition or therapy impacts participant well-being, confidence, anxiety, etc.

### 5 CONCLUSION

This paper provides an overview of the existing literature on XR games for mental health (XRMHGs) within the ISMAR, IEEE, and TVCG research communities. This review finds that XRMHGs extend beyond the simple gamification of therapeutic tasks and can be utilized to gain a deeper understanding of human behavior. This paper outlined those six games in detail to consolidate examples for future work. Moreover, this work offered six general considerations for future XRMHG researchers in our community based on prior existing XRMHG literature within our research space. 1) Researchers should be strategic and use their XRMHGs for more than surface level application evaluation, 2) game and XR elements should be well justified and tightly welded to both psychological theory and each other, 3) consider leveraging psychological foundations can manifest in several ways beyond simple gamification of therapeutic tasks, 4) carefully evaluate XRMHGs to protect (and actually measure with) vulnerable stakeholders, 5) consider using a workshop publication to academically describe the XRMHG in extreme detail for the academic record, and 6) be semantically precise when publishing future work in XRMHGs. In essence, the most important takeaways are to ensure that any XRMHG is a strong and considerate melding of XR, mental health, and games, to ensure that these facets are well explained in the literature, and provide meaningful evaluation of the XRMHG.

## 5.1 Limitations and Future Work

The main limitation of this scoping review is that it is limited to the ISMAR, IEEEVR, and TVCG publication venues. While limiting to a specific research community can be meaningful for outlining unique gaps and strengths within that community, at this time, this scoping review is of unknown generalizability outside of our research community. Future work should expand this review into a full systematic review with a broader set of databases.